\documentclass[12pt]{article}

\usepackage{newtxtext,newtxmath}
\usepackage{graphicx}

\usepackage[letterpaper,margin=1in]{geometry}

\renewenvironment{abstract}
	{\quotation}
	{\endquotation}

\date{}

\makeatletter
\renewcommand{\fnum@figure}{\textbf{Figure \thefigure}}
\renewcommand{\fnum@table}{\textbf{Table \thetable}}
\makeatother

\usepackage{scicite}
\usepackage{listings}
\usepackage{url}

\usepackage{booktabs}
\usepackage{xcolor}

\def\scititle{
    Scientific judgment drifts over time in AI ideation
}
\title{\bfseries \boldmath \scititle}

\author{
Lingyu Zhang$^{1}$, Mitchell Wang$^{2}$, Boyuan Chen$^{1, 2, 3\ast}$\\
\normalsize{$^{1}$Department of Electrical and Computer Engineering, Duke University}\\
\normalsize{$^{2}$Department of Computer Science, Duke University}\\
\normalsize{$^{3}$Department of Mechanical Engineering and Materials Science, Duke University}\\
\normalsize{$^\ast$To whom correspondence should be addressed; E-mail: boyuan.chen@duke.edu.}
}

\begin{document} 

% Insert the title and author list
\maketitle

\begin{abstract} \bfseries \boldmath
Scientific discovery begins with ideas, yet evaluating early-stage research concepts is a subtle and subjective human judgment. As large language models (LLMs) are increasingly tasked with generating scientific hypotheses, most systems implicitly assume that scientists’ evaluations form a fixed gold standard, assuming that scientists' judgments do not change. Here we challenge this assumption. In a two-wave study with 7,938 ratings from 63 active researchers across six scientific departments, each participant repeatedly evaluated a constant ``control'' research idea alongside AI-generated ideas. We find that expert evaluations are not stable: test--retest reliability of overall quality is only moderate (ICC$\approx 0.59$--$0.74$), indicating substantial within-participant variability even for identical ideas. Yet the internal structure of judgment remained stable, such as the relative importance placed on originality, feasibility, clarity, and other criteria. We then aligned an LLM-based ideation system to first-wave human ratings and used it to select new ideas. Although alignment improved agreement with Wave-1 evaluations, its apparent gains disappeared once drift in human standards was accounted for. Thus, tuning to a fixed human snapshot produced improvements that were transient rather than persistent. These findings reveal that human evaluation of scientific ideas is not static but a dynamic process with stable priorities and requires shifting calibration. Treating one-time human ratings as immutable ground truth risks overstating progress in AI-assisted ideation and obscuring the challenge of co-evolving with changing expert standards. Drift-aware evaluation protocols and longitudinal benchmarks may therefore be essential for building AI systems that reliably augment, rather than overfit to, human scientific judgment.

\end{abstract}

\noindent
%%%%%%%%%%%%%%%% INTRODUCTION %%%%%%%%%%%%%%% Average length for science papers: 800-1200 words

Early scientific-discovery systems relied on prespecified hypothesis templates \cite{dendral, mycin, eurisko, am} or curated knowledge bases \cite{wang2018abstract, wang2019paperrobot, he2020paperant}, which limited scale and adaptability.  Recent advances in Large Language Models (LLMs) \cite{gpt-3, gpt-4, gemini, gemini2.5, llama, llama2, llama3} have opened new possibilities. Trained on vast corpora, these models internalize extensive scientific and cultural knowledge and can be prompted in natural language to reason or hypothesize without domain-specific retraining. As a result, LLM-driven scientific ideation has rapidly emerged as a new frontier \cite{ai_scientist, canllms, scimon, ai_scientist_v2, ifargan2025autonomous, open_domain_discovery, hu2025nova, park2024can}.

Evaluating the quality of generated ideas remains a central obstacle. Implementing each idea and assessing them by outcomes is infeasible due to the resource- and time-intensive nature of scientific work. A practical alternative is to have scientists rate their potential value. However, designing evaluations to reflect genuine scientific value is difficult. Some studies use LLMs as proxies for human evaluators, either calibrated on human ratings \cite{researchagent} or justified by their correlation with human scores \cite{ai_scientist, ideabench}, while others rely entirely on expert judgment to guide iteration \cite{open_domain_discovery, scimon, canllms}. In all cases, human ratings are treated as a fixed ``gold standard'' for evaluation. Ideation systems are optimized to match that snapshot, assuming that scientists' criteria are stable and the resulting improvements will persist over time.

That assumption is rarely questioned. In practice, the perception of what counts as a good idea changes with shifts in culture, resources, and the direction of scientific progress. At the community level, priorities evolve as new methods or crises redirect attention. At the individual level, judgments shift as researchers gain experience, encounter new information, or simply revisit earlier impressions. The variability of human judgment over time is well documented \cite{kahneman2021noise}. Preferences drift in recommender systems as users adapt to the very content they consume \cite{koren2009collaborative, gama2014survey, kang2018self,carroll2022estimating}. Even the same annotator can assign different scores to identical text samples when asked twice \cite{kayapinar2014measuring}. Exposure to AI-generated material further alters human standards and beliefs \cite{rlhf_mislead, targeted_deception, glickman2025human}. Such variability grows with task difficulty \cite{abercrombie2023consistency, abercrombie2023temporal}.

Early-stage research ideas are often abstract and open-ended. Evaluating them is precisely this type of difficult task. Ratings must consider originality, feasibility, and impact, but the perceived value of these dimensions can change as contexts and norms evolve. This presents a critical concern. If an ideation system is tuned to ratings from one moment in time, the resulting ``improvements'' may vanish once evaluative standards move. Cross-study comparisons can also be misleading to reflect changes in human baselines rather than genuine methodological gains. 

The risk echoes Goodhart's law: a measure ceases to be good once it becomes a target \mbox{\cite{goodhart1984problems}}. Such dynamics arise in LLM training. Models must balance factual accuracy and user agreement. When optimization focuses narrowly on feedback collected at a single time point, it can overshoot and amplify the tendency rewarded most strongly at that time. Recent examples include ``sycophancy'' in GPT-4o \mbox{\cite{openai_sycophancy_gpt4o_2025, openai_expanding_sycophancy_2025}}, where models learned to agree excessively with users after alignment updates based on static preference data, followed by subsequent over-corrections once the issue was recognized \mbox{\cite{freedman_day_chatgpt_went_cold_2025}}. In AI, optimizing against an unmodeled, time-dependent human metric has produced fragile and unintended outcomes.

The temporal dynamics of scientists’ own evaluation criteria have not been systematically examined in ideation research. Here we ask: do scientists’ evaluations of research ideas change over time? And if so, in what ways? We examine two axes: (i) leniency or strictness, captured by shifts of absolute ratings across overall quality and specific value dimensions (e.g., originality, implementability); and (ii) relative importance, reflected by how scientists weight those dimensions when forming overall judgments. We study these questions in a two-wave study with a repeated control idea rated by the same participants. Finally, we ask whether tuning an ideation system to a fixed snapshot of human criteria yields improvements that persist once evaluation standards drift.

\begin{figure}[t!] % Do not use \begin{figure*}[ht]
	\centering
    \includegraphics[width=\textwidth]{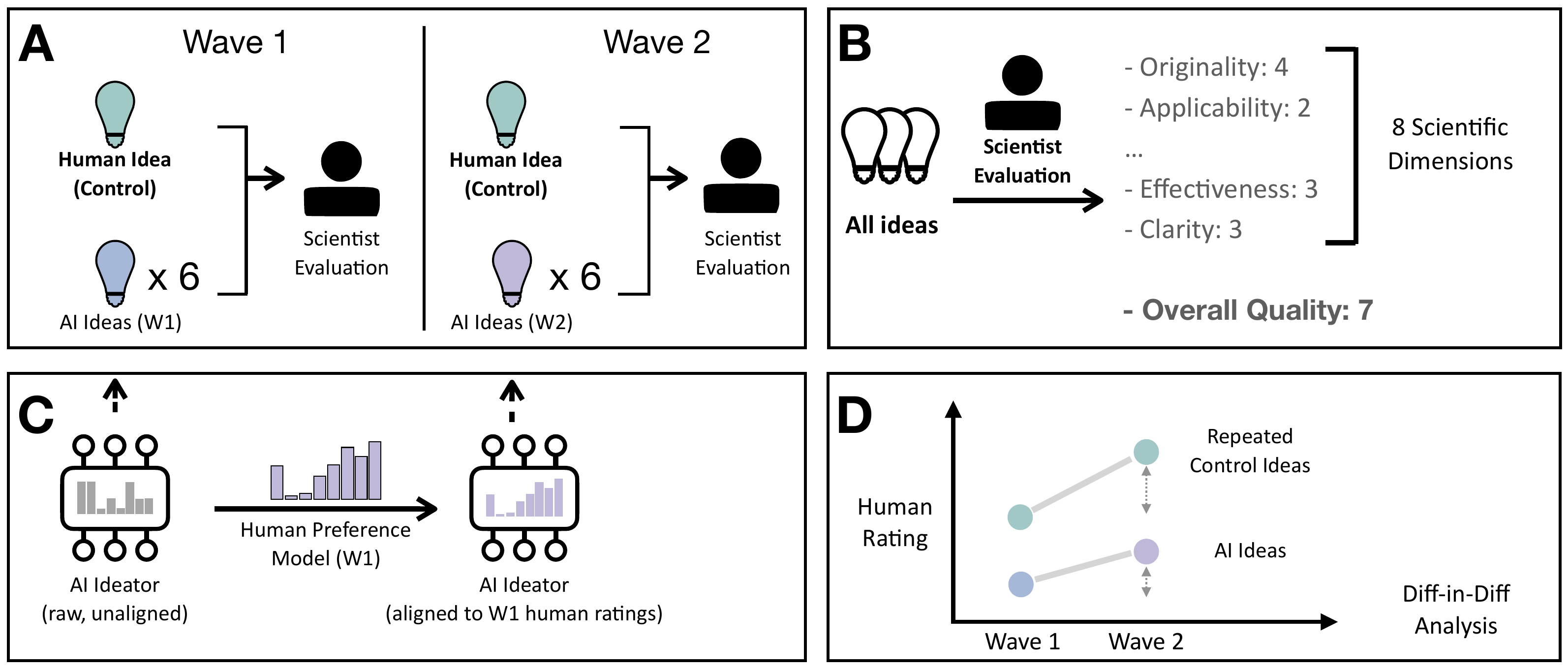}
	\caption{\textbf{Experimental design: two-wave evaluation with repeated controls and drift-aware analysis.}
    (A) Each scientist completed two evaluation waves. In each wave, they rated six AI-generated research ideas and one repeated human-written idea (control). The repeated control provides a direct measure of test–retest reliability and temporal drift in idea evaluation.
    (B) For every idea, participants rated eight scientific-value dimensions (e.g., originality, clarity, effectiveness) and an Overall Quality score (0–10 scale). We inferred the implicit weight assigned to each dimension by regressing Overall Quality on the dimension scores.
    (C) After Wave 1, human ratings were used to align an LLM-based evaluator to participants' preferences. This aligned evaluator then selected the six highest-ranked AI ideas for each scientist in Wave 2.
    (D) To assess whether alignment produced true improvement rather than overfitting to a transient snapshot of human criteria, we compared Wave 2 and Wave 1 ratings of AI ideas both naively and using a drift-corrected difference-in-differences anchored on the repeated control idea.}
	\label{fig:method} % give each figure a logical label name
\end{figure}

\section*{Study Overview}
We conducted a two-wave, within-participant study (Fig.~\ref{fig:method}).

In each wave, every scientist rated six AI-generated ideas and one repeated human-written control idea (Fig.~\ref{fig:method}A). The repeated control idea provided a direct measure of temporal drift and test-retest reliability in ratings. For each idea, participants rated eight scientific value dimensions adapted from Dean et al. \cite{idea_eval} and an Overall Quality score (0-10; 0=poor, 10=great; Fig.~\ref{fig:method}B). To model how scientists weighted dimensions when forming overall judgments, we regressed Overall Quality on the eight dimension scores. We tested whether these weights changed via Dimension $\times$ Wave interactions in a mixed-effects model. Between waves, we tuned the ideation system with Wave 1 human ratings. We then used the aligned evaluator to select Wave 2 AI ideas (Fig.~\ref{fig:method}C). 

% This allowed us to examine the consequences of modeling human judgment as a static criterion. 
We asked whether Wave 2 AI ideas were perceived as superior by the participants while controlling for drift using the unchanged control idea (Fig.~\ref{fig:method}D). This design enabled three analyses: (i) quantifying temporal changes in dimension-level perceptions and overall leniency, (ii) characterizing stability in cross-dimension weights, and (iii) testing whether tuning to a fixed human snapshot produces lasting gains while controlling for human judgment drift.

\subsection*{Participants}
63 active researchers from six departments at Duke University participated (21 faculty; 42 postdoctoral fellows, graduate students, or research staff; demographics and selection criteria in SM~\ref{st: participants}). Topics were customized to each participant’s domain to ensure subject familiarity (Methods~\ref{st: generator}). Participants were unaware that one idea would be repeated across waves. The final sample comprised of $N=63$ participants, yielding a total of 7,938 individual ratings (63 participants $\times$ 2 waves $\times$ 7 ideas $\times$ 9 scales). Sensitivity analyses across outlier removal methods are reported
in SM~\ref{st: screening}.

\subsection*{Temporal stability of ratings for the same idea}
\label{sec: dim change}
Participants rated each idea on eight dimensions: Originality, Paradigm Relatedness, Acceptability, Implementability, Applicability, Effectiveness, Implicational Explicitness, and Clarity; all on ordinal scales (SM~\ref{st: eval prompt}). Constructs and scales followed Dean et al. \cite{idea_eval}. Participants also rated on Overall Quality (0–10 scale).

To test whether perceptions of ideas changed over time, we implemented a test-retest design. Each participant re-evaluated the same human-written control idea in both waves. For each participant, the control idea was derived from a recent paper closely related to their field (excluding their own work; selection criteria in SM~\ref{st: control idea}). We shuffled the appearing orders of the ideas in all surveys. Depending on the response time, inter-wave intervals ranged from days to weeks (median [IQR] = 23 [9-35]; SM~\ref{st: time gap}).

We tested mean change of the control idea in Overall Quality with a paired $t$-test and visualized agreement using a Bland–Altman plot. Dimension-level shifts were evaluated with Wilcoxon signed-rank tests and Benjamini–Hochberg false-discovery-rate (FDR) correction across eight comparisons. Test–retest reliability was estimated using intraclass correlation coefficient ICC(A,1) for Overall Quality and quadratic weighted kappa for each dimension.

There can be several possible factors that influenced researchers' ratings between waves, such as familiarity effect, time gap, and exposure to new literature. While explaining the drift is not the focus of our study, we hypothesized and tested for a few variables that may be associated with the drift, listed in Tab.~\ref{tab:factors}. Notably we attempted to disentangle experiment-related artifacts and externally induced genuine criteria shifts. To do so, we constructed a variable, S-wave, to capture participants' own rubric applied to the same idea at each wave. Before rating ideas, participants reported how important each dimension was (0-10) when considering an idea. We normalized these importance weights to sum to 1. We then rescaled the eight dimensions' ratings to 0-1 and computed S-wave as the dot product of the normalized self-reported weights and the rescaled dimension ratings. Intuitively, S-wave summarizes participants' perception of an idea under their own rubric, which removes context and position effects. For details on the predictors' constructions, see Methods \ref{st: factors}. Using these predictors, we fit a mixed-effects model to the control idea's Overall Quality with participant random intercepts. 

\begin{table}[ht]
\centering
\caption{Fixed-effects of Overall Quality rating}
\label{tab:factors}
\begin{tabular}{lr}
\hline
\textbf{Fixed Effect Name} & Interpretation \\
\hline
Position & Position of the control idea within the wave \\
Context & Mean quality of ideas preceding the control in that wave  \\
Time gap & Days between waves  \\
Time of day & Time of day when surveys was submitted \\
Expertise & Faculty/Non-faculty \\
S-wave & Summary of self-reported rubric \\

\hline
\end{tabular}
\end{table}

\subsection*{Temporal stability of relative importance across dimensions}

We next tested whether the relative importance of the eight dimensions shifted. Verbal reports of evaluation criteria often diverge from the implicit rules guiding decisions \cite{nisbett1977telling}. To infer the latent weighting of each dimension, we modeled the Overall Quality ($Q_{ij}$) evaluated by participant $i$ for idea $j$ as a linear combination of the eight dimension scores across all ideas (one control and six AI ideas per participant) and both waves:\begin{equation}
Q_{ij} = \beta_0 + u_i + \beta_W \text{Wave}_{ij} + \sum_{k=1}^{8} \beta_k X_{ijk} + \sum_{k=1}^{8} \beta_{Wk} (\text{Wave}_{ij} \times X_{ijk}) + \epsilon_{ij}
\end{equation}
where $\beta_0$ is the fixed intercept, $u_i$ the random intercept for participant $i$, $\text{Wave}_{ij}$ the wave indicator and $X_{ijk}$ the score of the $k$-th scientific value dimension. The coefficients $\beta_k$ capture the main effect (relative weight) of each dimension while the interaction terms $\beta_{Wk}$ test for temporal shifts in those weights across waves. Finally, $\epsilon_{ij}$ represents the residual error. To enable direct comparability of coefficients, all eight dimension scores were transformed into standard scores within their respective waves.

\subsection*{Ideation-system performance under criteria shift}
\label{st: update}
Finally, we studied whether improvements achieved by tuning an ideation system to a fixed snapshot of human criteria persisted once evaluation standards changed. Wave 1 and Wave 2 each contained six AI-generated ideas per participant, selected from the same underlying candidate pool of 100 ideas using different evaluator criteria (heuristic weights in Wave 1; Wave-1-human-aligned weights in Wave 2). The two sets were therefore generally distinct but could overlap in principle. Across all participants, 2.78 out of 6 ideas appeared in both waves on average. Overlap was not unexpected: high-quality ideas might rank highly under both heuristic criteria and under aligned criteria. We treated the two sets as separate outputs of their respective selection processes and reported a robustness check restricted to non-overlapping ideas in Results.
The system implemented a representative LLM-based ideation pipeline aligned to Wave 1 human ratings. Wave 2 AI ideas were then selected by this aligned evaluator and rated by the same participants. We tested whether Wave 2 AI idea ratings improved over Wave 1 after accounting for concurrent shifts in human standards, as quantified by the control idea's change in rating. Next, we detail the core components of this study.

\textbf{Capturing a snapshot of human criteria.} LLM-driven idea generators are typically designed by grounding new ideas in prior literature \cite{ai_scientist, canllms, scimon, ai_scientist_v2, open_domain_discovery, hu2025nova, park2024can}. Following this principle, our system synthesized a pool of 100 candidate ideas per participant from structured summaries of their own publications (Methods~\ref{st: idea}). Since LLM-generated ideas vary in quality, previous systems often apply an automated evaluation step to score, refine or filter ideas \cite{canllms}. We again adopted this common practice by asking an LLM evaluator to rate each candidate idea using the same eight dimensions and prompts as the human raters. The Overall Quality score was computed as a weighted sum of dimensions. Without calibration, such LLM evaluators can diverge from human judgment both in per-dimension ratings and in inter-dimensional weighting. In Wave 1, we deployed the default evaluator with heuristic weighting to rank ideas and selected the top six for human evaluation (Methods~\ref{st: generator}). Prompts and LLM model specifics are provided in SM~\ref{st: generation prompt} and ~\ref{st: eval prompt}.

\textbf{Aligning system evaluation criteria with human snapshot.} After collecting the Wave 1 human ratings, we aligned the LLM evaluator to human scores in two steps: (i) fitting per-dimension linear calibrations mapping LLM scores to human scores and (ii) trained a ridge regression from the calibrated dimensions to the human-rated Overall Quality (train / validation split by idea; Methods~\ref{st: calibration}). The fitted model represented a fixed snapshot of Wave 1 human criteria. For each participant, this aligned evaluator re-scored the same candidate pool and selected the six highest-rated ideas for Wave 2 evaluation. We validated the calibration in two ways. We compared the residual error on a validation set of ideas with a uniform weights model. 

\textbf{Assessing genuine improvements of the system after alignment.} If human criteria were stationary, Wave 2 AI ideas should receive higher human ratings than Wave 1 AI ideas. In existing work, it is common to conclude the study once such result is obtained. However, if evaluative standards drift, such apparent improvements might vanish once concurrent changes in human criteria are taken into account. We therefore tested whether Wave 2 AI ideas received higher human ratings after drift correction. To achieve drift correction, we used two methods. First, we applied a standard mixed-effects model to estimate Wave 2 and Wave 1 ratings change while anchoring on the control ideas. However, model-based estimates can be opaque, so we also used a simple participant-level check: for each participant, we computed the change in their Control idea Overall Quality rating from Wave 1 to Wave 2 and subtracted that amount from their Wave 2 ratings of the AI ideas; we then compared these drift-corrected Wave 2 ratings to their Wave 1 ratings.

\subsection*{Summary of experimental logic}
Our two-wave design jointly measures temporal variability in scientists’ assessments of specific value dimensions, changes in the relative importance assigned to those dimensions, and the downstream impact of this variability on system improvement. The repeated control idea provides an anchor for test–retest reliability and for estimating temporal drift. Wave 1 captures a snapshot of human criteria. Aligning the ideation system to this snapshot yields a second set of AI ideas. Wave 2 evaluations, combined with the anchor, reveal whether perceived gains in AI-generated ideas persist once human judgment drift is explicitly modeled.

\section*{Results}

\subsection*{Scientists' ratings of the same idea change over time} 

\begin{figure}[t!] % Do not use \begin{figure*}
	\centering
    \includegraphics[width=0.9\textwidth]{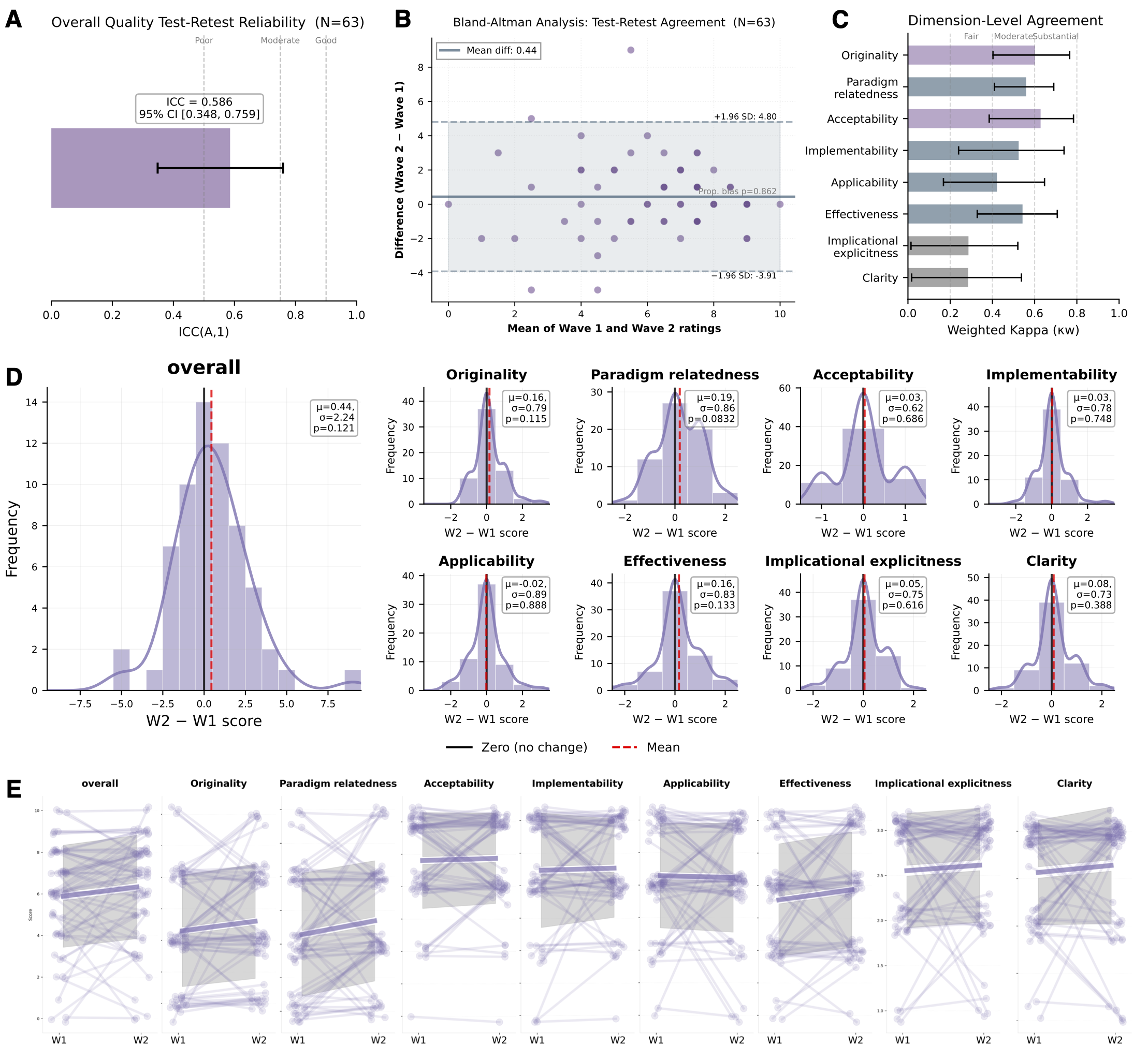}
    % for an image file named example_figure.* % Pick an appriopriate width for the size of the image % Captions go below figures
    \caption{\small\textbf{Scientists’ evaluations of the same scientific idea drift over time.}
    (A) Test-retest reliability of Overall Quality score for the repeated human-written control idea was moderate, indicating non-trivial variability even for identical stimuli.
    (B) Bland-Altman analysis of Overall Quality shows a positive mean shift between waves with no evidence of proportional bias.
    (C) Test-retest reliability of 8 scientific value dimensions (quadratic weighted kappa with 95\% CI).
    (D) Distribution of within-participant changes  (Wave~2 – Wave~1) for the control idea. Mean change was positive for Overall Quality and all dimensions except Applicability; statistical significance varied with outlier handling (see sensitivity analysis in SM~\ref{st: screening}). 
    (E) Paired trajectories for each participant across both waves illustrate consistent upward movement in Overall Quality and small positive shifts across most dimensions.} 
	\label{fig: change} 
\end{figure}

\textbf{Moderate reliability with a positive mean trend:} When the same participant re-rated the same control idea in Wave 2, the Overall Quality score increased by an average of 0.44 points on a 0-10 scale (paired $t$ test, N = 63 participants,  95\% CI $[-0.12,\;1.01]$, P = 0.121, dz = 0.198; Fig.~\ref{fig: change}D). Sensitivity analysis confirms consistently positive mean shift  (range $+0.31$ to $+0.48$ points), although its statistical significance depended on outlier handling method ( Table~\ref{tab:sensitivity} in SM~\ref{st: screening}). A Bland-Altman analysis showed positive bias and no evidence of proportional bias, i.e. higher-rated ideas did not drift differently from low-rated ones (slope = -0.008, P = 0.862, Fig.~\ref{fig: change}B).

At the dimension level, all scientific value dimensions except Applicability (six four-point scales and two three-point scales) shifted slightly upward, although none reached statistical significance in Wilcoxon signed-rank tests (all FDR-adjusted Ps $\geq$ 0.37; Table~\ref{tab:wilcoxon}). These results indicate a small positive mean trend in perceived idea quality, distributed broadly across dimensions rather than concentrated in any single criterion (Fig.~\ref{fig: change}E).

\textbf{Test-retest reliability:} Overall Quality ratings showed moderate test-retest reliability (ICC(A,1) = 0.586, 95\%~CI~[0.348, 0.759]; Fig.~\ref{fig: change}A). The Standard Error of Measurement (SEM) (1.585 points) yielded a Minimal Detectable Change of 4.394 points at 95\% confidence (MDC95), suggesting that only individual shifts larger than $\sim$4.4 points can be confidently interpreted as true perceptual changes. Note that even when individual measurements are noisy, aggregating within-person differences across many participants can yield a reliable estimate of the group mean trend. Reliability across dimensions, measured by quadratic weighted kappa, was similarly moderate and heterogeneous: Originality  0.60 [0.40, 0.77], Implementability 0.52 [0.24, 0.74], Acceptability 0.63 [0.38, 0.78], Effectiveness 0.54 [0.33, 0.71], Applicability 0.42 [0.17, 0.65], Paradigm relatedness 0.56 [0.41, 0.69], with lower agreement for Implicational explicitness 0.29 [0.01, 0.52] and Clarity 0.28 [0.02, 0.54] (Fig.~\ref{fig: change}C). These results indicate that instability is not limited to the Overall Quality score, but extends across the component dimensions used to evaluate scientific ideas.

\textbf{Drivers of the shift:} We modeled the control idea’s Overall Quality with participant random intercepts and fixed effects for Wave, Position, Context, Time gap, Time of day, Expertise, and S-wave (participants' rubric-weighted perception), with pre-specified Wave $\times$ Context, Wave $\times$ Time gap, and Wave $\times$ S-wave interactions (Table~\ref{tab:factors}). Results are shown in Fig.~\ref{fig: drivers}A. Wave
coefficient was positive but not significant ($\beta_{\textrm{wave}}$ = 0.144, SE = 0.204, z = 0.706, P = 0.480), consistent with the positive but non-significant trend from the paired $t$ test. S-wave strongly predicted Overall Quality ($\beta_\textrm{S-wave}$ = 1.839, SE = 0.209, z = 8.778, P $<$ 0.001), indicating that participants’ stated evaluation rubrics tracked their actual judgments. Position, Context, Time gap, Time of day, and Expertise were not significant (Ps $\geq$ 0.159). Interactions with Wave were not significant (Ps $\geq$ 0.289; full results see Tab.~\ref{tab: drivers}).

These predictors therefore do not explain the observed variability in repeated ratings. Several mechanisms remain plausible. One possibility is procedural familiarity: by Wave~2, participants may better understand the eight dimensions, reducing cognitive load and subtly changing how they score ideas. Our exploratory ablation analysis provides limited support for this explanation (SM~\ref{st: ablation}). A second possibility is cross-wave contrast or adaptation: as participants are exposed to different AI-generated ideas over time, their implicit baseline for judging the repeated human-written control idea may shift. Note that within-wave contextual contrast (the mean rating of preceding ideas) was not a significant predictor in our model, but cross-wave adaptation cannot be fully isolated with our design.

Finally, broader changes in scientific perspective arising from reflection, reading, or evolving field norms may also contribute. Our study was designed primarily to detect and quantify the consistency of human rating. We leave further exploration of drift mechanisms for future work. Regardless of mechanism, the practical implication of our study remains consistent: one-time human ratings are a moving target and should not be treated as immutable ground truth for iterative alignment.

\subsection*{Relative importance of dimensions remains stable over time} 
\begin{figure}[t] % Do not use \begin{figure*}
	\centering
    \includegraphics[width=\textwidth]{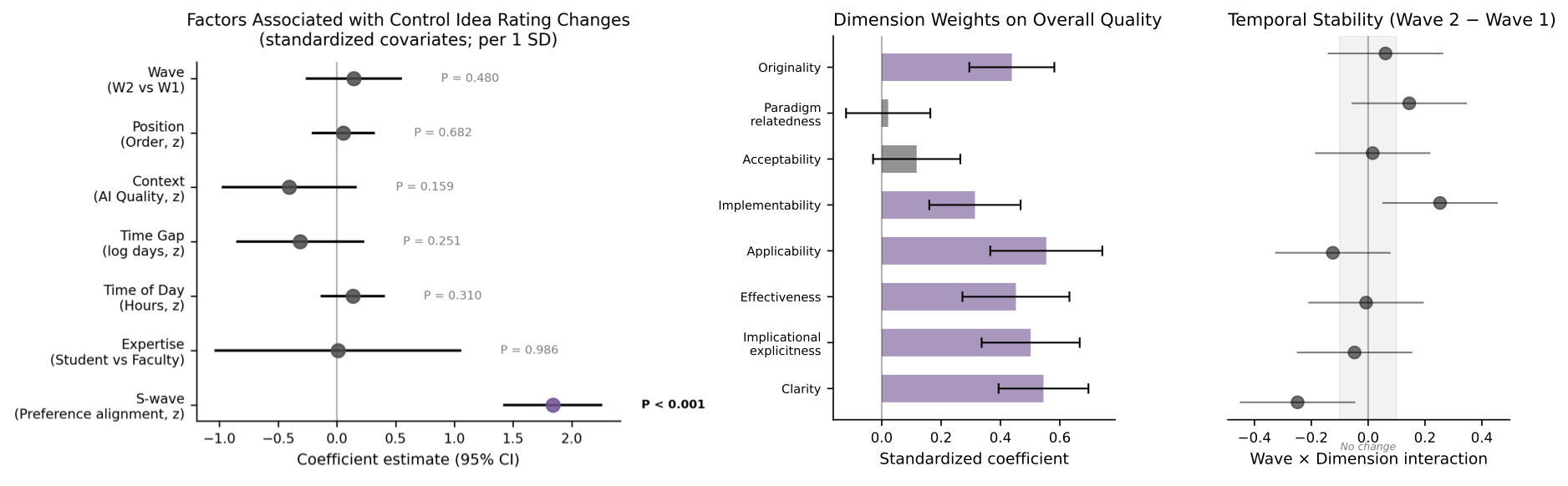}
	\caption{\textbf{Determinants and structure of scientific judgment across time.}
    (A) Mixed-effects regression examining factors associated with changes in ratings of the unchanged control idea. Self-reported evaluation weights (S-wave) strongly predicted ratings, confirming that participants' stated rubrics align with their actual judgments. Other factors, including position, contextual comparison, time gap, time of day, and expertise, are not significant.
    (B) Mixed-effects model of Overall Quality predicted by 8 dimensions, revealing the hierarchy of criteria driving scientific evaluation.
    (C) Wave $\times$ Dimension interaction terms reveal no significant changes in dimension weights over time, indicating that while absolute ratings drifted, the underlying structure of scientific judgment remained stable.}
	\label{fig: drivers}
\end{figure}

We next asked whether scientists altered the relative importance assigned to each of the eight dimensions. Across all ideas, we fit a mixed-effects model specified in Eq.~(1). Results are shown in Fig.~\ref{fig: drivers}B. Group-level coefficients showed strong positive associations for Applicability ($\beta$ = 0.554, SE = 0.096, z = 5.758, P $<$0.001), Clarity ($\beta$ = 0.545, SE = 0.077, z = 7.059, P $<$ 0.001), Implicational explicitness ($\beta$ = 0.502, SE = 0.084, z = 5.953, P $<$ 0.001), Effectiveness ($\beta$ = 0.452, SE = 0.092, z = 4.917, P $<$ 0.001), Originality ($\beta$ = 0.438, SE = 0.073, z = 5.999, P $<$ 0.001), and Implementability ($\beta$ = 0.314, SE = 0.079, z = 4.000, P $<$ 0.001), and non-significant association for Acceptability ($\beta$ = 0.118, P = 0.116) and Paradigm relatedness ($\beta$ = 0.022, P = 0.763).

Critically, we find no significant evidence that the weights applied to dimensions changed across waves (FDR-adjusted Ps $\geq$ 0.071 for all Dimension x Wave interaction terms; Fig.~\ref{fig: drivers}C). Hence, while repeated evaluations of the same idea exhibit variability over time overall, the cognitive weighting of dimensions of what makes a ``good idea'' remained stable over the study interval.

\subsection*{Aligning to a fixed criterion results in only transient gains}

\textbf{Validating the alignment:} Aligning the LLM evaluator to Wave 1 human ratings sharply reduced prediction error on a held-out validation set (1.86 vs. 15.35 of uniform weights' residual; Fig.~\ref{fig:calibrator}A), capturing a fixed snapshot of human criteria at Wave 1.

\begin{figure}[t!] % Do not use \begin{figure*}
	\centering
    \includegraphics[width=0.95\textwidth]{waves_fig/fig3_revision.png}
	\caption{\textbf{Alignment to a snapshot of human judgment yields transient rather than persistent gains.}
    (A) Aligning the LLM evaluator to Wave 1 human ratings substantially reduced prediction error on both training and held-out validation ideas, compared with a uniform-weights baseline.
    (B) Raw human ratings initially suggested improvement: Wave 2 AI ideas scored 0.29 points higher than Wave 1 ideas. The repeated human control idea increased by 0.44 points over the same period, revealing concurrent upward drift.
    (C) Difference-in-differences after drift correction: no persistent gain beyond the control’s drift.}

	\label{fig:calibrator}
\end{figure}

\textbf{Snapshot gains do not generalize under drift.} While raw human ratings of newly selected Wave 2 AI ideas were 0.29 points higher than those of Wave 1 AI ideas, this improvement did not persist after adjusting for concurrent drift in the unchanged control idea. For each participant, we subtracted their own control idea change (Wave 2 $-$ Wave 1) from their AI idea change - a participant-level difference-in-differences that assumes each individual drifts in their own way. We find that the AI idea rating improvement in Wave 2 disappears after adjusting for concurrent drift (mean = $-0.156$; 95\%~CI $[-0.766,\;0.454]$; $d_z = -0.064$; $t = -0.511$; $P = 0.611$; Fig.~\ref{fig:calibrator}C). As a robustness check, restricting the analysis to non-overlapping AI ideas (those uniquely contributed by each selection process) yielded the same conclusion: a raw improvement of $+0.57$ points did not persist after drift correction ($+0.12$ points, $P = 0.703$). Thus, the apparent gains achieved by aligning to a fixed snapshot of human criteria did not persist once human judgment drift was accounted for, highlighting that human-in-the-loop tuning may result in transient rather than persistent improvements.

Across both waves, the human-written control ideas (derived from vetted published papers) received substantially higher Overall Quality ratings than the AI-generated ideas. We note that the control idea was deliberately selected from recent, cited publications and therefore represents a high bar.

\section*{Discussion}

Evaluating scientific ideas is uniquely challenging. It requires balancing originality, feasibility, potential impact, and many other value dimensions that are inherently uncertain and context-dependent. Our longitudinal study shows that these human evaluative criteria are not fixed in time.
Ratings for an unchanged control idea showed a small positive mean trend between waves, and test–retest reliability across dimensions was only moderate and heterogeneous indicating that repeated judgments of the same idea can vary substantially even within the same participant. When we tuned an AI evaluator to match a one-time snapshot of human criteria (Wave 1) and used it to select ideas for Wave 2, the apparent gains vanished once we corrected for concurrent drift using the repeated control. Optimizing to a fixed snapshot improved a transient metric but did not yield genuine and persistent improvements in human ratings.

These findings challenge a prevailing practice in the emerging literature on AI-assisted scientific ideation: that one-off human evaluations can be treated as fixed gold standards. Whether systems directly learn a human judgment model from a single wave of ratings or indirectly optimize it by running single-round human rating comparisons, both approaches assume that human criteria are stationary. Our results show that this assumption is fragile. As evaluative standards evolve, gains measured against a fixed snapshot of humans preferences may overstate true progress. When subsequent studies benchmark against these inflated baselines, the field risks compounding irreproducible improvements and slowing genuine advancement. Such risks are analogous to model drift in recommender systems or biasing to drifted human preference in human-guided AI systems \cite{zhang2024guide, jipref, sonright}.

At the same time, we found that the ``structure'' of human judgment remained largely stable. The relative importance scientists assigned to different dimensions, such as originality, clarity, or implementability, did not change significantly across waves. This suggests that while overall leniency and absolute ratings drifted, the hierarchy of values that underlies scientific idea evaluation persisted. Human evaluation, therefore, is not unreliable noise, but a dynamical system with stable internal structure and shifting baseline calibration. This distinction matters for how we design and interpret human-in-the-loop AI ideation systems. Models that assume stability risk overfitting to transient conditions, but models that explicitly track and adapt to evolving human criteria may bring genuine and persistent advancements over time.

The mechanisms underlying these changes remain an open question. Our analysis finds no evidence that within-session order effects, local contextual comparisons, time-of-day, or time gap explain the observed variability. Several other explanations remain plausible. Procedural familiarity, cross-wave contrast/adaptation, and deeper shifts in scientific values driven by exposure to newer information or an evolving field could be contributors. Disentangling these mechanisms will require richer experimental designs. For example, a multi-arm Wave~2 design including aligned, unaligned, and deliberately ``reverse-aligned'' ideas could more directly test whether improvements persist beyond concurrent baseline changes. Such designs must balance causal resolution with survey burden and potential fatigue effects.

Our finding aligns with lessons from adjacent fields. In recommender systems, temporal drift in user preferences is expected and routinely modeled \cite{koren2009collaborative, gama2014survey}. In behavioral and cognitive research, intra-rater variability is treated as an intrinsic property of human judgment rather than an error to be eliminated \cite{nesselroade2004intraindividual, vaughan2023within}. Scientific ideation systems can adopt similar strategies by incorporating temporal anchors, repeated controls, and adaptive evaluators.

\subsection*{Practical implications for evaluation of AI ideation systems}

Our findings suggest an actionable template for evaluating human-in-the-loop AI ideation systems under nonstationary criteria. Here we list what we believe to be the most important points to keep in mind while developing future AI ideation systems.

\noindent\textbf{Report stability metrics.} Any measurement of human responses that assumes stable human ratings should include test-retest reliability analyses. Report standard reliability metrics such as ICC and kappa to quantify practical significance and stability.

\noindent\textbf{Insert repeated controls.} As a baseline monitor of human criteria change, repeated ideas should be included in each iteration of the system update.

\noindent\textbf{Use multiple waves to test generalization.} Test whether performance improvements persist across evaluation rounds when the human reference shifts.

\noindent\textbf{Model drift explicitly.} Treat human criteria as a dynamic variable and include time and context-dependent predictors to model it, such as the time gap between measurements, the time of day, and context information surrounding the rated item.

We hope these practices can help create an evaluation framework that is both more reproducible and more reflective of how human judgment evolves in practice.

\subsection*{Limitations and future directions}

Our work represents an initial step. There are several future opportunities to improve the study. Our two-wave design demonstrates the existence of preference drift but does not map its longer-term dynamics. Future work spanning more waves could reveal whether evaluation criteria eventually stabilize, oscillate, or continue to drift over a longer period of time.

Our human evaluative model was a simple linear combination of dimensions; more expressive models such as dynamic latent-variable approaches may capture additional nuance of the human criteria such as nonlinear or interaction effects. We measured perceived scientific quality, but not yet downstream scientific impact. Understanding whether early-stage idea ratings predict discoveries, publications, or funded projects will require long-term field experiments and future advancements in automated scientific experiments.

Our study uses one repeated control idea per participant. While the 63 distinct control ideas provide robustness at the population level, multiple repeated controls per participant would improve estimation of within-person variability and help further disentangle procedural familiarity from substantive criteria change. We made a pragmatic tradeoff here: additional repeated items would lengthen an already 20-30 minute survey and could introduce fatigue effects. Future studies could consider shortening idea formats to keep survey length manageable.

Finally, though our study covers multiple departments and different level of researchers, our sample comprised scientists from a single institution. Cross-community replication could reveal whether drift patterns generalize across scientific cultures. Moreover, our results are based on a representative yet specific LLM-based ideation system and alignment method; future studies could explore how these findings generalize across different models and system designs. With more accumulated data from the community, more advanced alignment methods such as model fine-tuning could be implemented. Additionally, due to the consideration of obtaining higher response rates, we only included one repeated control idea per participant; more controls could further reduce variance. On the other hand, while repeated items is a standard way of measuring test-retest reliability, we cannot exclude familiarity effects as confounding variables that might have influenced ratings change. Future work could explore more sophisticated designs to address these risks.

%Inter-rater agreement.
\section*{Conclusion}
Human evaluation of scientific ideas is not fixed. In a two-wave study, we find that repeated ratings of the same ideas exhibit only moderate reliability, alongside small but variable changes in overall scores, while the relative importance assigned to different evaluation dimensions remains stable. Aligning an AI ideation system to a one-time snapshot of human judgment produced transient gains that vanished once drift was considered. These results call for longitudinal and drift-aware evaluation protocols that incorporate repeated controls, report stability metrics, and explicitly model the dynamics of human judgment. Recognizing that scientific value judgments are dynamic rather than static may be key to building AI systems that not only generate ideas, but continue to evolve with the scientists who assess, improve, and perhaps take inspiration from them.

%%%%%%%%%%%%%%%% REFERENCES %%%%%%%%%%%%%%%

\clearpage % Clear all remaining figures and tables then start a new page

\bibliography{science_template} % for a file named science_template.bib
\bibliographystyle{sciencemag}

%%%%%%%%%%%%%%%% ACKNOWLEDGEMENTS %%%%%%%%%%%%%%%

\section*{Acknowledgments}

\paragraph*{Funding:}
This work is supported by DARPA FoundSci program under award HR00112490372.

\paragraph*{Author contributions:}
Conceptualization: L.Z., B.C. Methodology: L.Z., M.W., and B.C. Data Collection: L.Z, M.W., and B.C. Project administration: B.C. Supervision: B.C. Writing - original draft: L.Z. and B.C. Writing - reviewing and editing: L.Z., M.W., and B.C.

\paragraph*{Competing interests:}
There are no competing interests to declare.

\paragraph*{Data and materials availability:}
Original Survey Responses, Processing and Analysis Code for all experiments are available at \url{https://github.com/generalroboticslab/IdeationEval}

%%%%%%%%%%%%%%%% SUPPLEMENT LIST %%%%%%%%%%%%%%%

% List the contents of your Supplementary Materials, including the numbers of any
% supplementary figures, tables, external data files etc. and any references that are
% cited only in the supplement. In this example, refs. 7-8 are cited only in the supplement.
% Fill out your numbers accordingly and delete any lines that aren't applicable.
\subsection*{Supplementary materials}
% Materials and Methods\\
Supplementary Text\\
Figs. S1 to S3\\
Tables S1 to S4\\
% References \textit{(7-\arabic{enumiv})}\\ % automatically fills out the last reference number
% (filling out the other numbers automatically is possible but fiddly and liable to break)
% Movie S1\\
% Data S1

%%%%%%%%%%%%%%%% END OF MAIN TEXT %%%%%%%%%%%%%%%

\newpage

%%%%%%%%%%%%%%%% START OF SUPPLEMENT %%%%%%%%%%%%%%%

% Figures, tables, equations and pages in the supplement are numbered S1, S2 etc.
\renewcommand{\thefigure}{S\arabic{figure}}
\renewcommand{\thetable}{S\arabic{table}}
\renewcommand{\theequation}{S\arabic{equation}}
\renewcommand{\thepage}{S\arabic{page}}
\setcounter{figure}{0}
\setcounter{table}{0}
\setcounter{equation}{0}
\setcounter{page}{1} % not 0 as \newpage already started a supplementary page
% References continue the numbering from the main text.

%%%%%%%%%%%%%%%% SUPPLEMENT TITLE PAGE %%%%%%%%%%%%%%%

\begin{center}
\section*{Supplementary Materials for\\ \scititle}

% Author list for the supplement
% Indicate the corresponding authors, but do NOT include institutions here
% It would be nice if the template auto-generated this, but doing so is complicated...
% First~Author$^{\ast\dagger}$,
% A.~Scientist$^\dagger$,
% Someone~E.~Else\\ % we're not in a \author{} environment this time, so use \\ for a new line
% \small$^\ast$Corresponding author. Email: example@mail.com\\
% \small$^\dagger$These authors contributed equally to this work.

\author{
Lingyu Zhang$^{1}$, Mitchell Wang$^{2}$, Boyuan Chen$^{1, 2, 3\ast}$\\
\normalsize{$^{1}$Department of Electrical and Computer Engineering, Duke University}\\
\normalsize{$^{2}$Department of Computer Science, Duke University}\\
\normalsize{$^{3}$Department of Mechanical Engineering and Materials Science, Duke University}\\
\normalsize{$^\ast$To whom correspondence should be addressed; E-mail: boyuan.chen@duke.edu.}
}

\end{center}

% Fill out the numbers for each type of supplementary material,
% and delete any lines that aren't applicable.
% These are just example numbers that don't match the rest of this template.
\subsubsection*{This PDF file includes:}
% Materials and Methods\\
Supplementary Text\\
% Figures S1 to S3\\
% Tables S1 to S4\\
% Captions for Movies S1 to S2\\
% Captions for Data S1 to S2
Figs. S1 to S3\\
Tables S1 to S4\\

% \subsubsection*{Other Supplementary Materials for this manuscript:}
% Movies S1 to S2\\
% Data S1 to S2

\newpage

%%%%%%%%%%%%%%%% MATERIALS AND METHODS %%%%%%%%%%%%%%%
% \section*{Materials and Methods}
\section*{Supplementary Text}

\section*{Overview}

We conducted a two-wave within-participant study to quantify temporal variability in scientists' evaluations of research ideas and its implications for AI-driven ideation. Each participant rated a set of AI-generated ideas and one repeated human-written control idea across two sessions. The first wave provided a snapshot of human evaluative criteria, and the second wave tested whether those criteria and model performance tuned to them remained stable. All participants provided informed consent prior to the experiment. The study protocol was approved by the Duke University Institutional Review Board.

\section{Participants}
\label{st: participants}
Participants were recruited from the following six departments at Duke University: Biomedical Engineering, Computer Science, Electrical and Computer Engineering, Mechanical Engineering and Materials Science, Psychology and Neuroscience, and Sociology. 

\textbf{Faculty (seed) recruitment and filtering.} To identify active faculty researchers and to obtain a publication set for customizing survey topics, we filtered faculty candidates by whether they had at least one first- or last-author paper published since January 2020 with external citations. This filter was used to target active labs and to support publication-based customization; it was not intended as a hard eligibility constraint for non-faculty participants.

\textbf{Non-faculty recruitment and rationale.} Non-faculty participants (postdoctoral fellows, graduate students, and research staff) were recruited from the corresponding faculty participants' labs or field-adjacent labs as active researchers in the same topical area. Because non-faculty researchers may not yet have first/last-author publications meeting the faculty filter, we did not apply that criterion to non-faculty participation. To avoid introducing additional topic variability, non-faculty participants evaluated the same domain-customized survey content as their advisor or most matching faculty participant.

For the first wave of our study, we received 41 responses from faculty members. We used these response ratings to conduct the calibration step that yielded the Wave 2 candidate idea pools. Wave 2 surveys are only sent to researchers who responded to the Wave 1 survey. Researchers who responded to both waves were our final participants, comprising 21 faculty members and 42 non-faculty researchers.

\section{Idea evaluation survey}
\label{st: survey}
For each wave, each participant completed a customized Google Forms survey lasting approximately 20 to 30 minutes. In each survey, participants evaluated seven ideas: six AI-generated ideas and one repeated human-written control idea (presented in random order). For each idea, participants rated eight scientific value dimensions: Originality, Paradigm relatedness, Acceptability, Implementability, Applicability, Effectiveness, Implicational explicitness, and Clarity, along with an Overall Quality score (0-10 scale). Detailed instructions and prompts are provided in SM~\ref{st: eval prompt}.

\section{Structured idea representations}
\label{st: idea}

To ensure control and AI-generated ideas were indistinguishable in style and structure, all ideas (control and AI-generated) were formatted into a structured representation:

\begin{lstlisting} 
    "Title: ,
    Tldr: ,
    Problem: ,
    Solution: ,
    Key Insight: ,
    Inspiration: ,
    Figure Caption: ,
    Proposed Idea: "
    
\end{lstlisting}

When extracting ideas from existing papers, the pdf file was first converted to text using \texttt{pymupdf4llm} \cite{pymupdf4llm}. Then the idea was extracted using an LLM model (\texttt{gpt-4o-mini}) with the prompt provided in SM~\ref{st: idea extraction prompt}. The temperature was set to 0.2 and top\_p to 0.1. 
For both extracting and generating ideas, we used OpenAI's structured output API to ensure the desired format is enforced.

\section{AI ideation system}

\subsection{Generator design}
\label{st: generator}

For every participant, we generated a pool of 100 customized ideas with an LLM-based ideation system. In Wave 1, six ideas were selected based on a heuristic criterion (SM~\ref{st: calibration}). In Wave 2, six ideas were reselected based on a human-aligned criterion.

\subsubsection{Inputs}
High-impact science often arises from recombining distant knowledge domains \cite{shi2023surprising,youn2015invention} and atypical mixtures of conventional and novel elements \cite{uzzi2013atypical}. Inspired by this, our idea generator took in two input ideas: (i) a ``conventional'' idea that reflects a participant’s recent work, and (ii) a ``novel'' idea that is highly cited and potentially disruptive, providing a source of inspiration. For each participant, we formed two pairing modes per wave:  cross-domain pairs and in-domain pairs. In both modes, the ``conventional'' idea was the participant's most recent first-/last-author research article with at least one external citation (no self-citations). In a cross-domain pair, the inspiration-providing ``novel'' idea was sampled from a highly-cited paper from a different department. In an in-domain pair, the ``novel'' idea was sampled from the same department. Survey, review, and perspective articles were excluded from both sets.

It is not often the case that inspiration from an impactful paper can be directly applied to any given conventional problem, so we did not assume all ideas generated with this method will be of high quality. As a recent study has suggested \cite{canllms}, the advantage of LLM-driven ideation is to generate large quantities of ideas, and the most useful ones may emerge at the top of them. A total of 50 cross-domain pairs and 50 in-domain pairs were sampled for each participant to form the full 100 candidate idea pool. In each wave, the top 3 cross-domain and the top 3 in-domain ideas were selected for inclusion in the surveys.

\subsubsection{Domain-compatibility pre-screening} 
\begin{figure}[ht] % Do not use \begin{figure*}[h]
	\centering
	\includegraphics[width=0.85\textwidth]{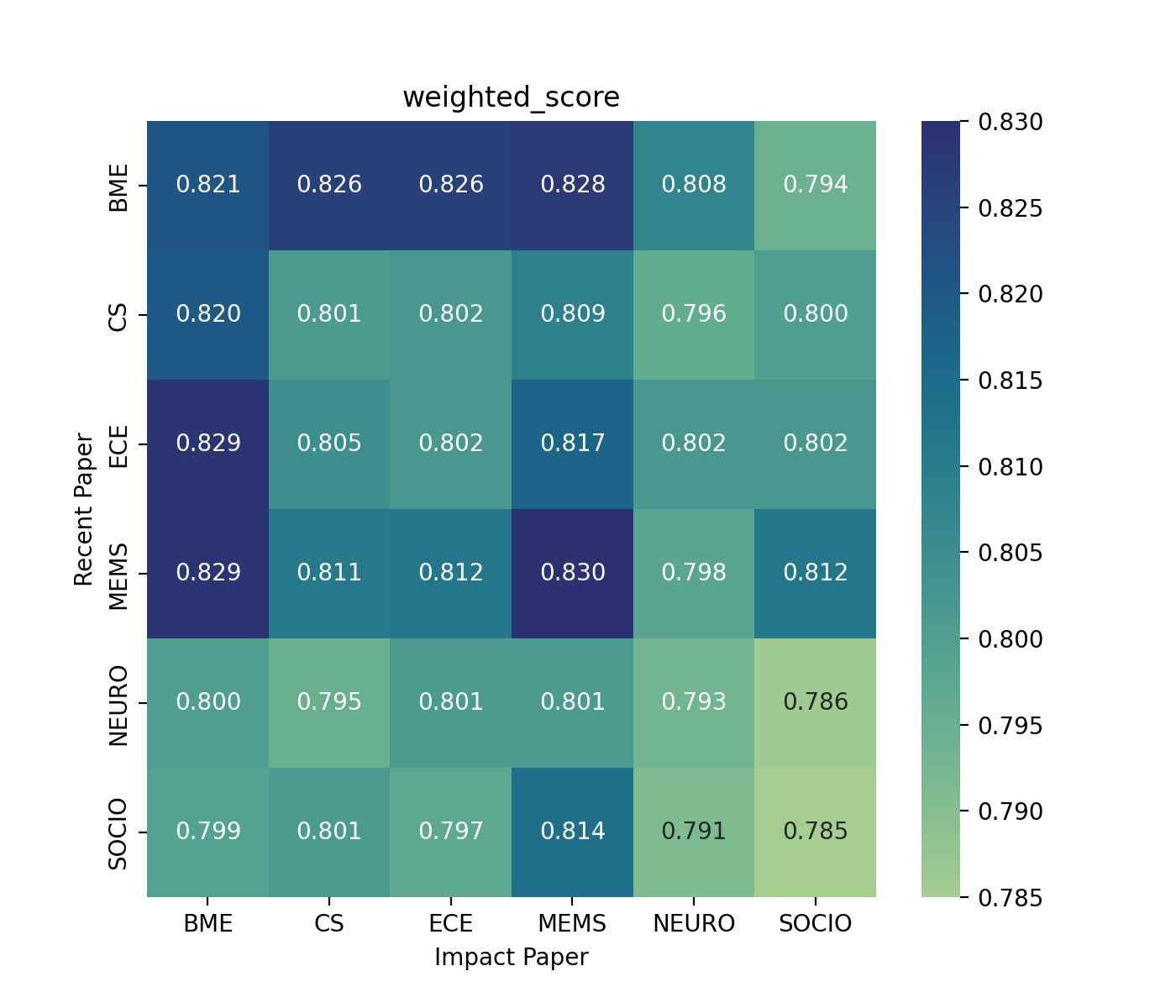}
	\caption{\textbf{Department Compatibility Matrix}}
	\label{fig: compatibility} % give each figure a logical label name
\end{figure}

Because not all cross-department pairings were equally promising, we performed a preliminary compatibility screen. For each department pair (e.g. [Biomedical Engineering, Sociology]), we sampled one conventional paper and ten candidate novel papers. For each of the ten pairs, we prompted the generator five times, yielding 50 ideas. We then scored these idea with our heuristic LLM evaluator (SM~\ref{st: evaluator}). The final compatibility of a pair of departments was calculated by the average LLM evaluator rating. Compatibility matrix is shown in Fig.~\ref{fig: compatibility}

\subsubsection{Model and prompting}
We used OpenAI’s \texttt{gpt-4o-2024-08-06} as the generator. For idea generation we kept temperature and top\_p at their defaults (1.0) to encourage diversity. We used OpenAI's structured output API to ensure that ideas were returned in a standardized structure SM~\ref{st: idea}. To avoid trivial re-statements of known work, the prompt explicitly instructed the model to be transformative, concrete, and not similar to existing works. The full generation prompt can be found in SM~\ref{st: generation prompt}.

\subsection{Evaluator design}
\label{st: evaluator}

Each generated idea was evaluated by an automated LLM-based evaluator that rated the same eight dimensions as human participants. To ground evaluations in existing literature, we implemented a retrieval-augmented pipeline. A \texttt{gpt-4o-mini} model generated five keyword summaries per idea, which were submitted to Semantic Scholar with five such queries. The retrieved papers' titles, tldrs, and abstracts were saved into a pool of potentially relevant papers. We then filtered this pool of papers by their embedding similarity to the idea to be evaluated and kept those with a similarity higher than 0.88. We used OpenAI's \texttt{text-embedding-large} model for this step. The threshold was decided after manual inspection for relevancy on a small set of examples.

The evaluation prompt (SM~\ref{st: eval prompt}) used the same instructions and dimension anchors as those shown to human raters. For more consistent evaluation, we follow best practices in using LLMs as evaluators and compute a probability-weighted summation of the output token scores as the final score \cite{liu2023g-eval, lee2024fleur, huang2024instupr}.

% alignment with ICLR evals
% The two main sections of the supplement can be split up using headings.
% \textbf{Cross-domain pairing} Results

\section{Human-written control idea selection}
\label{st: control idea}

Each participant was assigned a repeated control idea derived from a real published research paper. The paper topic was close to their research field, but not authored by them. This control idea appeared once in each survey wave. 

To identify suitable control papers, we compiled each participant's publication list from Google Scholar using their Scholar ID. We retrieved their most recent paper that had a citation that was ``Highly Influenced'' by it, as defined by the Semantic Scholar API. We restricted the papers to be non-review and non-survey articles published after January 2020. In this way, we sampled papers closely related to the participant's research while minimizing recognition effects. 

For consistency across experimental conditions, the control idea was converted into a standardized text summary (SM~\ref{st: idea}) using the same processing pipeline applied to AI-generated ideas (including PDF-to-text conversion, key idea extraction, and formatting).

\section{Predictors of rating change}
\label{st: factors}
\noindent\textbf{Position.} The index of the control idea in the survey (1-7).

\noindent\textbf{Context.} Humans are subject to context effects, where judgments may be shaped by surrounding information within the session. We calculated the mean quality of ideas preceding the control to capture this effect.

\noindent\textbf{Quality.} Quality is approximated by an LLM-evaluator to remove temporal variability. To handle cases where the control idea appeared first in the survey, we centered the mean quality of preceding ideas by the mean quality of all AI-generated ideas. By doing so, if no AI ideas appeared before the control idea, the value of the Context predictor would simply be zero.

\noindent\textbf{Time gap.} Since the time gap distribution was skewed (see SM~\ref{st: time gap}), we took the log value of the days between responses.

\noindent\textbf{Time of day.} Time of day the survey was submitted, centered by noon time (12 pm).

\noindent\textbf{Expertise.} Indicator variable of whether the participant is a faculty member.

\noindent\textbf{S-wave.} A composite variable reflecting the participant's self-reported criteria of scientific ideas. To obtain this, we included a question at the beginning of every survey, asking the participant to rate (0-10) how important each of the 8 scientific value dimension was when considering the overall quality of a scientific idea. We then normalized them to have a sum of 1. Using this set of self-reported importance weights, we weighed their ratings along each dimension of the control idea and calculated the S-wave variable as the sum. While this approach could not fully eliminate confounding factors (such as familiarity effect on participants' self-reported criteria), it represented our best effort to disentangle external influences from methodological artifacts. 

\section{Alignment of ideation system}
\label{st: calibration}

Before aligning with human Wave 1 ratings, the evaluator was assigned with fixed heuristic weights to each dimension:
% The initial weights over the 8 dimensions are chosen by the following heuristic:
\begin{lstlisting}
    "originality": 0.25
    "paradigm_relatedness": 0.25
    "acceptability": 0.05
    "implementability": 0.10
    "applicability": 0.10
    "effectiveness": 0.15
    "implicational_explicitness": 0.05
    "clarity": 0.05
\end{lstlisting}

After collecting Wave 1 human ratings, we used the 41 faculty members' ratings to align the evaluator using a two-step alignment process. First, per-dimension linear models mapped LLM evaluator scores to human ratings. Second, we trained a ridge regression model to predict human Overall Quality score from the eight aligned dimension ratings. A 20\% validation split was used to perform a grid search for the ridge regularization parameter. The resulting coefficients defined the updated evaluator weights used for selecting Wave 2 AI ideas.

\section{Data quality and outlier screening.} 
\label{st: screening}
\begin{figure}[ht]
\centering
\includegraphics[width=0.9\linewidth]{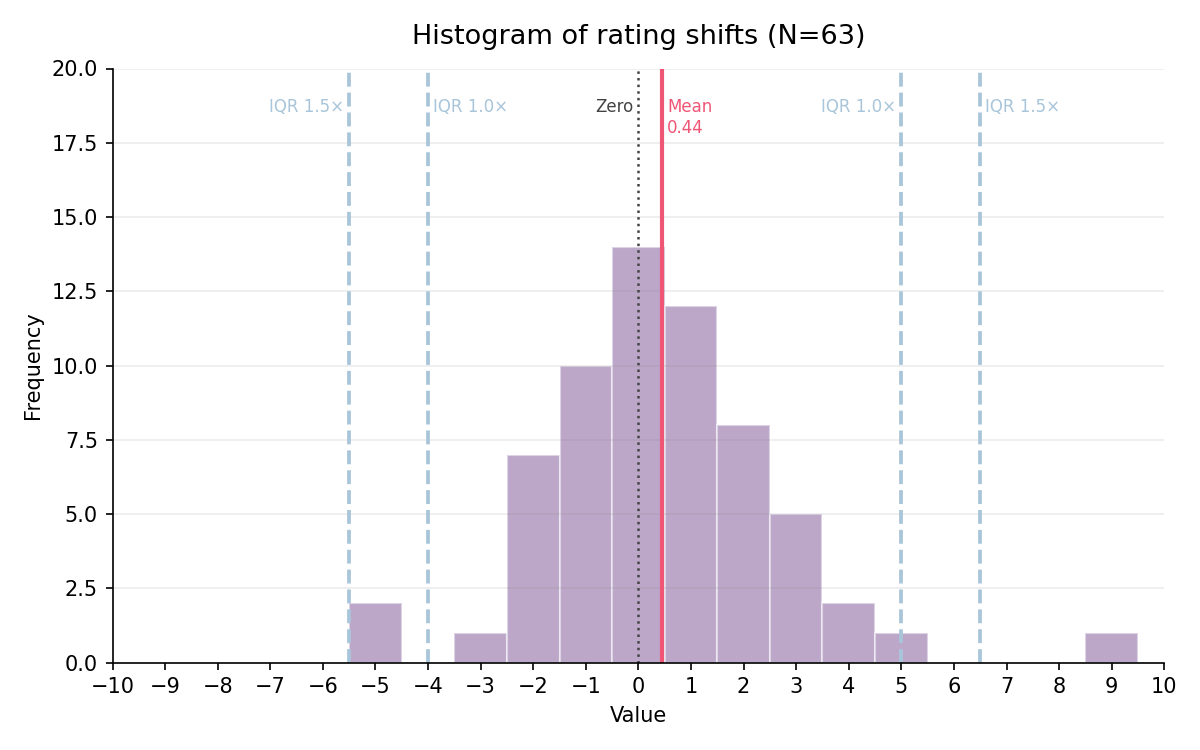}
\caption{Full distribution of W1-W2 repeated idea drift values with IQR fences.}
\label{fig:outlier_histogram}
\end{figure}

To isolate changes in human evaluative criteria (as opposed to differences in idea content), longitudinal stability tests focused on the unchanged human control idea. We conducted a sensitivity analysis across eight analytical variants, spanning no outlier removal through four standard methods - Tukey IQR with multipliers 1.0$\times$, 1.5$\times$, and 3.0$\times$; standard $z$-score cutoffs ($z \geq 2.0$, $z \geq 2.5$, $z \geq 3.0$); and MAD-based modified $z$-score ($\geq 3.5$) - applied to the full sample of $N=63$. Full distribution and IQR fences is visualized in Fig.~\ref{fig:outlier_histogram}. Key results across all variants are presented in Table~\ref{tab:sensitivity}. The mean overall quality drift showed a consistent positive trend (range $+0.31$ to $+0.48$ points across variants), with statistical significance depending on the aggressiveness of outlier removal ($p$ range 0.036--0.226). Test-retest reliability (ICC) and the difference-in-differences conclusion were consistent across all variants. The primary analysis reported in the main text uses the full sample with no outlier removal ($N=63$).

\section{Exploratory Study on Procedural Familiarity}
\label{st: ablation}
To investigate whether procedural familiarity explains the drift, a separate $n=24$ non-faculty participants completed a fully-repeated condition. In this ablation group, all 7 AI plus human ideas in Wave 1 and Wave 2 are repeated. Across $24\times7=168$ fully-repeated idea pairs, ratings showed a small and non-significant change between waves ($\Delta=+0.13$, 95\% CI $[-0.24,\;0.51]$, $p=0.478$). This provides exploratory evidence that the survey procedure itself (repeated exposure to the same format) does not drive any systematic rating increase on its own. 

\section{Time Gap Between Responses}
\label{st: time gap}

\begin{figure}[ht] % Do not use \begin{figure*}[h]
	\centering
    \includegraphics[width=0.7\textwidth]{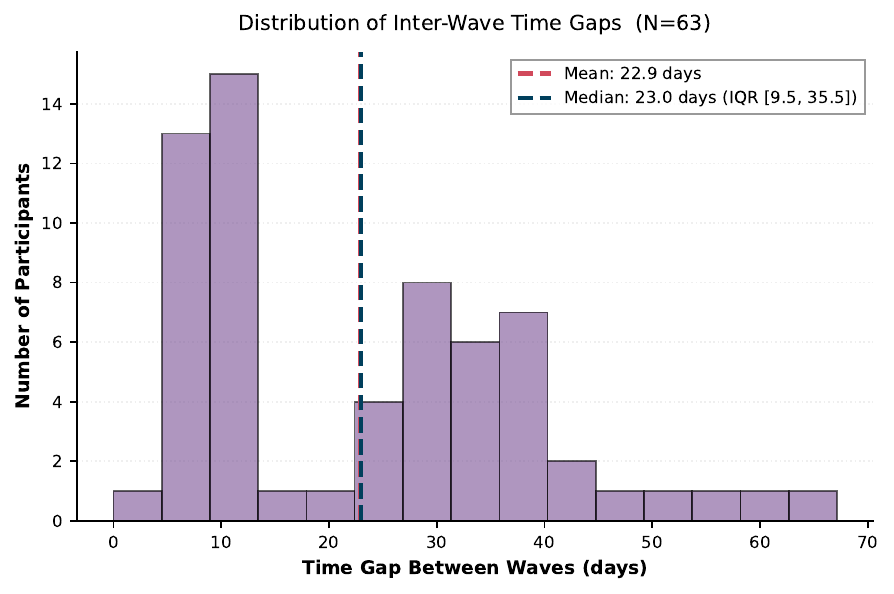}
	\caption{\textbf{Time gap between waves}}
	\label{fig:time gap} % give each figure a logical label name
\end{figure}

Information can be found in Fig.~\ref{fig:time gap}

\section{Literature Search Query Generation Prompt}
\label{st: query prompt}

\begin{lstlisting}
SYSTEM PROMPT: "You are a proficient researcher."
USER PROMPT: f"Your task is to come up with 5 text queries with no more than 8 words each to search on a scholarly search engine for papers related to the following idea:
    {idea}
    The queries must be as concise as possible while being as informative about the key innovation of the idea as possible. The queries should also be diverse in phrasing to maximize the chances of finding relevant papers."
\end{lstlisting}

\section{Idea Extraction Prompt}
\label{st: idea extraction prompt}

\begin{lstlisting}
SYSTEM PROMPT: "You are tasked with retrieving specific information from a research paper. Try to use the authors' own words when possible, and be as detailed as possible."
USER PROMPT: f"""{paper_text}.
    From the above paper, retrieve the following information:
    1) The tile of the paper.
    2) The TLDR of the paper.
    3) The specific research problem the paper is trying to address, in the context of existing literature.
    4) The specific and unique solution that the paper proposes.
    5) The key insight or observation that led to the solution.
    6) The inspiration of the paper.
    7) The caption of the first figure in the paper. 
    8) A summary of the specific proposed idea of the paper.
    Be as detailed as possible when extracting the above, and do not exclude any information that is relevant to the above points. """
\end{lstlisting}

\section{Idea Generation Prompt}
\label{st: generation prompt}
\begin{lstlisting} 
You are a visionary researcher. Your task is to synthesize insights from summaries of past academic papers, some of them conventional, some of them seem unrelated but may be inspiring, and generate a novel research idea that is groundbreaking, original, feasible, and concrete. The idea should have substantial impact and is inspired by at least one of the papers (not necessarily all of them). Identify structures, patterns or implicit insights from the inspiring papers that could be potentially connected and applied to conventional ideas. When presenting the proposed idea, format it the same way as past ideas without mentioning the past papers.

Example creative ideas (for demonstration, not in the final format):
    1. Evolutionary Algorithms: Natural Selection selects the fittest individuals for reproduction, leading to the development of evolutionary algorithms that mimic this process to solve optimization problems.
    2. Sonar Technology: Bats and dolphins use echolocation that produces sound waves and listens to the echoes to navigate and locate objects, inspiring the development of sonar technology.
    3. Neural Networks: The human brain processes information through interconnected neurons, leading to the development of artificial neural networks that mimic this structure to solve complex problems.

    Key objectives:
        1. Propose transformative improvements or entirely new directions.
        2. Ensure ideas are detailed, concrete, specific, feasible and actionable.
        3. Make sure the ideas are well-conveyed and easy to understand.
        4. Avoid any similarity with the related works.
        {conventional_papers}
        {similar_works}
        {novel_papers}
        
\end{lstlisting}

\section{Evaluation Instructions/Prompt}
We used the same instructions for both human and automated LLM evaluation. The eight dimensions are introduced by Dean et al. \cite{idea_eval}. The original anchors were developed in the context of the restaurant business and tourism ideas. To suit our study better, we prompted an LLM (OpenAI o1 model) to derive science-related prompts for each anchor, using the original anchors as prompts. Below are the final evaluation instructions/prompts.

\label{st: eval prompt}
\begin{lstlisting}
Given the relevant literature above, evaluate an idea proposed by a PhD student along the following dimensions. For each dimension, provide a score between (1-4) or (1-3) for implicational_explicitness and clarity, following the scales below.

originality: The degree to which the idea is not only rare but is also ingenious, imaginative, or surprising.
    4: Not expressed before (rare, unusual) and Ingenious, imaginative or surprising
        Examples: 
        - Develop a method to record and interpret human thoughts directly into digital format.
        - Propose a theory that unifies quantum mechanics and general relativity through multidimensional time.
    3: Unusual, interesting; shows some imagination
        Examples:
        - Investigate the use of genetically modified fungi to decompose plastic waste rapidly.
        - Explore the potential of harnessing lightning as a renewable energy source.
    2: Interesting
        Examples:
        - Study the effects of intermittent fasting on gut microbiota diversity.
        - Analyze the impact of urban green spaces on mental health in city residents.
    1: Common, mundane, boring
        - Measure the growth rate of common bacteria under different temperature conditions.
        - Conduct a survey on students' study habits during exam periods.

paradigm_relatedness: The degree to which an idea preserves or modifies a paradigm. Paradigm-Modifying ideas are sometimes radical or transformational.
    4: Paradigm Breaking. For scientific research, a paradigm breaking idea introduces new elements and changes the fundamental concepts and experimental practices.
        Examples:
        - Propose that consciousness is a fundamental component of the universe and develop experiments to test this.
        - Suggest that the laws of physics vary across different regions of the cosmos and devise methods to detect such variations.
    3: Paradigm Stretching. For scientific research, a paradigm stretching idea changes the fundamental concepts and experimental practices.
        Examples:
        - Use quantum computing to simulate complex biological processes at the molecular level.
        - Apply principles from neuroscience to develop artificial intelligence that mimics human cognition.
    2: Slightly Paradigm Stretching. For scientific research, a slightly paradigm stretching idea introduces new elements.
        Examples:
        - Integrate virtual reality technology into psychological therapy for treating phobias.
        - Utilize drone technology for real-time environmental monitoring.
    1: Paradigm Preserving. For scientific research, conduct research with usual methods and concepts.
        Examples:
        - Conduct experiments using established chemical reactions to synthesize known compounds.
        - Perform standard statistical analysis on existing data sets to confirm previous findings.

acceptability: The degree to which the idea is socially, legally, or politically acceptable.
    4: Common strategies that violate no norms or sensibilities.
        Examples:
        - Publish collaborative research findings in open-access journals.
        - Attend international conferences to present research and network.
    3: Somewhat uncommon or unusual strategies that don't offend sensibilities.
        Examples:
        - Create an interactive online platform to engage the public in research projects.
        - Use gamification techniques to teach complex scientific concepts.
    2: Offends sensibilities somewhat but is not totally unacceptable.
        Examples:
        - Conduct experiments involving the genetic modification of animal embryos.
        - Use controversial data sources acquired under questionable circumstances.
    1: Radically violates laws or sensibilities or Totally unacceptable business practice. 
        Examples:
        - Perform human experimentation without informed consent.
        - Falsify data to achieve desired research outcomes.

implementability: The degree to which the idea can be easily implemented.
    4: Easy to implement at low cost or non-radical changes.
        Examples:
        - Switch to digital data recording to reduce paper waste.
        - Use existing equipment during off-peak hours to maximize usage.
    3: Some changes or reasonably feasible promotions or events.
        Examples:
        - Apply for additional funding to expand the scope of the current project.
        - Train staff to use new software that improves data analysis efficiency.
    2: Significant change or expensive or difficult but not totally impossible to implement.
        Examples:
        - Upgrade laboratory facilities with state-of-the-art equipment.
        - Organize a multidisciplinary research team requiring coordination across departments.
    1: Totally infeasible to implement or extremely financially nonviable.
        Examples:
        - Construct a research facility on the moon to conduct low-gravity experiments.
        - Build a particle accelerator the size of the solar system to test new physics theories.

applicability: The degree to which the idea clearly applies to the stated problem.
    4: Solves an identified problem that is directly related to the stated problem (do X to get Y, and Y is part of the stated problem) 
        Examples:
        - Develop a rapid diagnostic test for a widespread infectious disease.
        - Create a biodegradable alternative to single-use plastics to reduce pollution.
    3: Solves an implied problem that is related to the stated problem (do X to get an implied Y, which applies to the stated problem) 
        Examples:
        - Implement machine learning algorithms to improve data processing efficiency.
        - Introduce energy-efficient practices in the lab to reduce operational costs.
    2: May have some benefit within a special situation and somehow relates to the stated problem (do X, which somehow relates to the stated problem)
        Examples:
        - Organize community outreach programs to enhance public understanding of science.
        - Incorporate team-building exercises to improve lab morale.
    1: Intervention is not stated or does not produce a useful outcome (no X) or (do X for useless Y) 
        Examples:
        - Rearrange the lab furniture to improve aesthetics.
        - Change the lab coat color to boost team spirit.

effectiveness:  The degree to which the idea will solve the problem.
    4: Reasonable and will solve the stated problem without regard for workability (If you could do it, it would solve the main problem) 
        Examples:
        - Discover a universal cure for viral infections by targeting a common viral protein.
        - Invent a clean energy source that provides unlimited power without environmental impact.
    3: Reasonable and will contribute to the solution of the problem (It helps, but it is only a partial solution)
        Examples:
        - Develop a new drug that significantly improves treatment outcomes for a disease.
        - Design a water filtration system that removes 99% of contaminants.
    2: Unreasonable or unlikely to solve the problem (It probably will not work) 
        Examples:
        - Propose to cool the Earth's temperature by painting all rooftops white.
        - Attempt to increase human lifespan to 200 years by dietary changes alone.
    1: Solves an unrelated problem (It would not work, even if you could do it) 
        Examples:
        - Study butterfly wing patterns to improve internet security protocols.
        - Analyze ancient texts to solve modern traffic congestion issues.

implicational_explicitness: The degree to which there is a clear relationship between the recommended action and the expected outcome.
    3: Implication is clearly stated and makes sense (do X so that Y) 
        Examples:
        - Increase sample size to enhance the reliability of experimental results.
        - Implement quality control measures to reduce errors in data collection.
    2: Implication is not generally accepted or is vaguely stated (do X, which solves a not-generallyaccepted Y ) or (do X which solves a vaguely stated Y) 
        Examples:
        - Use homeopathy to treat bacterial infections, expecting improved patient outcomes.
        - Rely on psychic abilities to predict natural disasters.
    1: Implication is not stated, even though relevant (do X without a stated Y)
        Examples:
        - Collect soil samples from various locations.
        - Measure atmospheric pressure regularly.

clarity: The degree to which the idea is clearly communicated with regard to grammar and word usage.
    3: Crisp, with standard usage, including complete sentences or well-developed phrases, and every word is commonly understood 
        Examples:
        - Investigate the role of gut microbiota in regulating immune system function.
        - Examine the correlation between air pollution levels and respiratory health issues.
    2: Understandable, with acceptable usage or understandable phrases; some words might be known only within a small context; sentences might contain fragments or be incomplete (yet understandable)
        Examples:
        - Test how plants grow under different lights.
        - Look at effects of exercise on heart health.
    1: Vague or ambiguous words or use of poor language structure 
        Examples:
        - Do stuff in lab to see what happens.
        - Check out some science things to get results.
\end{lstlisting}

\newpage

\begin{table}[t]
\centering
\caption{Sensitivity analysis across analytical variants.
  \textbf{Removed \%}: Percentage of the sample removed by the outlier screening method.
  \textbf{ICC}: test-retest reliability of Overall Quality for the repeated control idea — the primary, robust finding.
  \textbf{Drift $\Delta$ / $p$}: mean change and paired $t$-test for the control idea between waves; positive values indicate higher ratings in Wave~2.
  \textbf{Raw AI $\Delta$}: mean change in AI idea ratings before drift correction.
  \textbf{DiD $p$}: participant-level difference-in-differences $p$-value testing whether AI idea improvement exceeds concurrent drift in control idea ratings. ICC and DiD conclusions are consistent across all variants.}
\vspace{10pt}
\label{tab:sensitivity}
\small
\begin{tabular}{lrrrrrrr}
\toprule
\textbf{Version} & $N$  & \textbf{Removed \%} & \textbf{ICC} &  \textbf{Drift $\Delta$}& \textbf{Drift $p$}  & \textbf{Raw AI $\Delta$} & \textbf{DiD $p$} \\
\midrule
no screening            & 63 &   0 & $0.586$ & $+0.44$ & $0.121$ & $+0.29$ & $0.611$ \\
IQR $3.0\times$        & 63 &   0 & $0.586$ & $+0.44$ & $0.121$ & $+0.29$ & $0.611$ \\
IQR $1.5\times$        & 62 & 1.6 & $0.671$ & $+0.31$ & $0.226$ & $+0.29$ & $0.954$ \\
IQR $1.0\times$        & 60 & 4.8 & $0.718$ & $+0.48$ & $0.036$ & $+0.34$ & $0.609$ \\
$z \geq 3.0$           & 62 & 1.6 & $0.671$ & $+0.31$ & $0.226$ & $+0.29$ & $0.954$ \\
$z \geq 2.5$           & 62 & 1.6 & $0.671$ & $+0.31$ & $0.226$ & $+0.29$ & $0.954$ \\
$z \geq 2.0$           & 59 & 6.3 & $0.738$ & $+0.41$ & $0.064$ & $+0.36$ & $0.845$ \\
MAD $\geq 3.5$         & 62 & 1.6 & $0.671$ & $+0.31$ & $0.226$ & $+0.29$ & $0.954$ \\
\bottomrule

\end{tabular}
\end{table}

\begin{table}[ht]
\centering
\caption{Test--retest reliability}
\label{tab:test_retest}
\begin{tabular}{lrrr}
\hline
\multicolumn{4}{l}{\textbf{Overall Reliability}} \\
% \multicolumn{4}{l}{ICC(A,1) = 0.721; SEM = 1.124; MDC$_{95}$ = 3.116} \\
\multicolumn{4}{l}{ICC(A,1) = 0.586; SEM = 1.585; MDC$_{95}$ = 4.394} \\
\hline
\textbf{Dimension} & \textbf{Kappa$_{qw}$} & \textbf{CI$_{lo}$} & \textbf{CI$_{hi}$} \\
\hline
Originality & 0.602 & 0.402 & 0.766 \\
Paradigm relatedness & 0.560 & 0.409 & 0.691 \\
Acceptability & 0.628 & 0.385 & 0.784 \\
Implementability & 0.524 & 0.240 & 0.739 \\
Applicability & 0.422 & 0.168 & 0.647 \\
Effectiveness & 0.543 & 0.328 & 0.707 \\
Implicational explicitness & 0.287 & 0.014 & 0.520 \\
Clarity & 0.285 & 0.017 & 0.538 \\
\hline
\end{tabular}
\end{table}

\begin{table}[ht]
\centering
\caption{Wilcoxon signed-rank test (R2$-$R1) per dimension with FDR correction}
\label{tab:wilcoxon}
\begin{tabular}{lrrrrrr}
\hline
\textbf{Dimension} & \textbf{HL} & \textbf{CI$_{lo}$} & \textbf{CI$_{hi}$} & \textbf{$p$} & \textbf{$p_{FDR}$} & \textbf{Reject} \\
\hline
Originality & 0.0 & 0.0 & 0.0 & 0.123 & 0.370 & False \\
Paradigm relatedness & 0.0 & 0.0 & 0.0 & 0.083 & 0.370 & False \\
Acceptability & 0.0 & 0.0 & 0.0 & 0.683 & 0.840 & False \\
Implementability & 0.0 & 0.0 & 0.0 & 0.840 & 0.840 & False \\
Applicability & 0.0 & 0.0 & 0.0 & 0.809 & 0.840 & False \\
Effectiveness & 0.0 & 0.0 & 0.0 & 0.139 & 0.370 & False \\
Implicational explicitness & 0.0 & 0.0 & 0.0 & 0.627 & 0.840 & False \\
Clarity & 0.0 & 0.0 & 0.0 & 0.383 & 0.767 & False \\
\hline
\end{tabular}
\end{table}

\begin{table}[ht]
\centering
\caption{Mixed model on control Overall (drivers of change; standardized covariates)}
\label{tab: drivers}
\begin{tabular}{lrrrrrr}
\hline
\multicolumn{7}{l}{\textbf{Mixed Linear Model Regression Results}} \\
\hline
& Coef. & Std.Err. & $z$ & $P>|z|$ & [0.025 & 0.975] \\
\hline 
Intercept & 6.045 & 0.411 & 14.723 & 0.000 & 5.240 & 6.849 \\
Expertise & 0.009 & 0.531 & 0.017 & 0.986 & $-1.031$ & 1.050 \\
Wave & 0.144 & 0.204 & 0.706 & 0.480 & $-0.256$ & 0.544 \\
Position & 0.054 & 0.132 & 0.410 & 0.682 & $-0.205$ & 0.313 \\
Context & $-0.406$ & 0.288 & $-1.408$ & 0.159 & $-0.970$ & 0.159 \\
Time Gap & $-0.313$ & 0.273 & $-1.147$ & 0.251 & $-0.848$ & 0.222 \\
Time of Day & 0.136 & 0.134 & 1.015 & 0.310 & $-0.126$ & 0.398 \\
S-wave & 1.839 & 0.209 & 8.778 & 0.000 & 1.428 & 2.249 \\
Wave$\times$Context & 0.350 & 0.330 & 1.061 & 0.289 & $-0.297$ & 0.997 \\
Wave$\times$Time Gap & $-0.137$ & 0.209 & $-0.658$ & 0.510 & $-0.546$ & 0.272 \\
Wave$\times$S-wave & $-0.109$ & 0.238 & $-0.458$ & 0.647 & $-0.576$ & 0.357 \\
\hline
Group Var & 1.349 & 0.471 & & & & \\
\hline
\end{tabular}
\end{table}

\newpage

\end{document}